\documentclass[aps,prl,preprint,tightenlines,superscriptaddress,showpacs,byrevtex]{revtex4}
\usepackage{graphicx} 
\usepackage{dcolumn}  

\begin{document}
\vspace*{-3\baselineskip}
\resizebox{!}{3cm}{\includegraphics{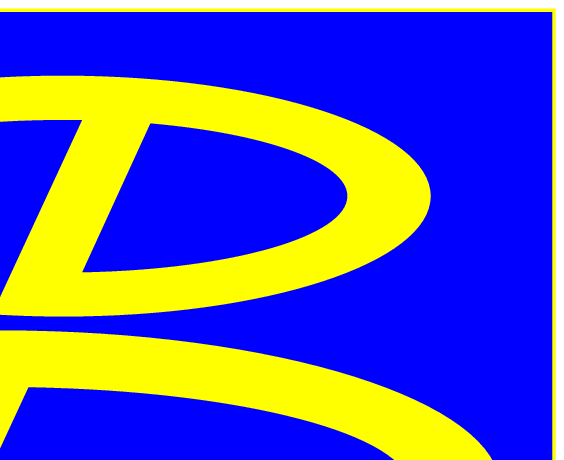}}

\preprint{\vbox{  \hbox{Belle Preprint 2004-11}
                 \hbox{KEK Preprint 2003-143}
}}

\title{ \quad\\[0.5cm] { 
           Study of CP Violating Effects in Time Dependent 
          $B^0(\bar{B^0}) \rightarrow D^{(*)\mp}\pi^{\pm}$ Decays}}


\affiliation{Budker Institute of Nuclear Physics, Novosibirsk}
\affiliation{Chiba University, Chiba}
\affiliation{University of Cincinnati, Cincinnati, Ohio 45221}
\affiliation{University of Frankfurt, Frankfurt}
\affiliation{Gyeongsang National University, Chinju}
\affiliation{University of Hawaii, Honolulu, Hawaii 96822}
\affiliation{High Energy Accelerator Research Organization (KEK), Tsukuba}
\affiliation{Hiroshima Institute of Technology, Hiroshima}
\affiliation{Institute of High Energy Physics, Chinese Academy of Sciences, Beijing}
\affiliation{Institute of High Energy Physics, Vienna}
\affiliation{Institute for Theoretical and Experimental Physics, Moscow}
\affiliation{J. Stefan Institute, Ljubljana}
\affiliation{Kanagawa University, Yokohama}
\affiliation{Korea University, Seoul}
\affiliation{Kyungpook National University, Taegu}
\affiliation{Swiss Federal Institute of Technology of Lausanne, EPFL, Lausanne}
\affiliation{University of Ljubljana, Ljubljana}
\affiliation{University of Maribor, Maribor}
\affiliation{University of Melbourne, Victoria}
\affiliation{Nagoya University, Nagoya}
\affiliation{Nara Women's University, Nara}
\affiliation{National United University, Miao Li}
\affiliation{Department of Physics, National Taiwan University, Taipei}
\affiliation{H. Niewodniczanski Institute of Nuclear Physics, Krakow}
\affiliation{Nihon Dental College, Niigata}
\affiliation{Niigata University, Niigata}
\affiliation{Osaka University, Osaka}
\affiliation{Panjab University, Chandigarh}
\affiliation{Peking University, Beijing}
\affiliation{Princeton University, Princeton, New Jersey 08545}
\affiliation{RIKEN BNL Research Center, Upton, New York 11973}
\affiliation{University of Science and Technology of China, Hefei}
\affiliation{Seoul National University, Seoul}
\affiliation{Sungkyunkwan University, Suwon}
\affiliation{University of Sydney, Sydney NSW}
\affiliation{Tata Institute of Fundamental Research, Bombay}
\affiliation{Toho University, Funabashi}
\affiliation{Tohoku Gakuin University, Tagajo}
\affiliation{Tohoku University, Sendai}
\affiliation{Department of Physics, University of Tokyo, Tokyo}
\affiliation{Tokyo Institute of Technology, Tokyo}
\affiliation{Tokyo Metropolitan University, Tokyo}
\affiliation{Tokyo University of Agriculture and Technology, Tokyo}
\affiliation{University of Tsukuba, Tsukuba}
\affiliation{Utkal University, Bhubaneswer}
\affiliation{Virginia Polytechnic Institute and State University, Blacksburg, Virginia 24061}
\affiliation{Yokkaichi University, Yokkaichi}
\affiliation{Yonsei University, Seoul}
  \author{T.~R.~Sarangi}\affiliation{High Energy Accelerator Research Organization (KEK), Tsukuba} 
  \author{K.~Abe}\affiliation{High Energy Accelerator Research Organization (KEK), Tsukuba} 
  \author{K.~Abe}\affiliation{Tohoku Gakuin University, Tagajo} 
  \author{T.~Abe}\affiliation{High Energy Accelerator Research Organization (KEK), Tsukuba} 
 \author{H.~Aihara}\affiliation{Department of Physics, University of Tokyo, Tokyo} 
  \author{Y.~Asano}\affiliation{University of Tsukuba, Tsukuba} 
  \author{V.~Aulchenko}\affiliation{Budker Institute of Nuclear Physics, Novosibirsk} 
  \author{T.~Aushev}\affiliation{Institute for Theoretical and Experimental Physics, Moscow} 
  \author{S.~Bahinipati}\affiliation{University of Cincinnati, Cincinnati, Ohio 45221} 
  \author{A.~M.~Bakich}\affiliation{University of Sydney, Sydney NSW} 
% ...
\collaboration{The Belle Collaboration}


\begin{abstract}
  We report measurements of time dependent decay rates for 
  $B^0(\bar{B}^0) \rightarrow D^{(*)\mp}\pi^{\pm}$ decays 
  and extraction of CP violation parameters containing $\phi_3$. 
  Using fully reconstructed $D^{(*)}\pi$ events from a $140\ {\rm fb}^{-1}$ 
  data sample collected at the $\Upsilon(4S)$ resonance,   
  we obtain the CP violation parameters for $D^* \pi$ and $D \pi$ decays,
  $2R_{D^{(*)} \pi} \sin (2\phi_1 + \phi_3 \pm \delta_{D^{(*)} \pi})$,  
  where $R_{D^{(*)} \pi}$ is the ratio of the magnitudes of the 
  doubly-Cabibbo-suppressed and Cabibbo-favoured amplitudes,
  and $\delta_{D^{(*)} \pi}$ is the strong phase difference between them.
  Under the assumption of $\delta_{D^{(*)} \pi}$ being close to 
  either 0 or $180^{\circ}$, we obtain 
  $|2R_{D^* \pi} \sin (2\phi_1 + \phi_3)| = 0.060 \pm 0.040(\mathrm{stat}) 
  \pm 0.019(\mathrm{sys})$ and 
  $|2R_{D \pi} \sin (2\phi_1 + \phi_3)| = 0.061 \pm 0.037(\mathrm{stat}) 
  \pm 0.018(\mathrm{sys})$. 
  
\end{abstract}

\maketitle
\tighten

The good agreement between direct measurements of 
$\sin 2\phi_1$~\cite{belle_sin2phi1,babar_sin2phi1}
and the outcome of global fits to the CKM quark mixing matrix elements
~\cite{PDG} strongly supports the 
standard model explanation of  CP violation.  
To determine whether it is the complete description or 
whether additional factors come into play, further 
measurements of other CKM parameters are required. 
Among these parameters  $\phi_3$ is of particular importance.  
The measurements of time-dependent decay rates of 
$B^0 (\bar{B}^0) \to D^{(*)\mp}\pi^{\pm}$ provide a theoretically clean 
method for extracting $\sin(2\phi_1+\phi_3)$, since loop diagrams do not 
contribute to these decays~\cite{dunietz}.   

\begin{figure}[bht]
\begin{minipage}{4.0cm}
\begin{center}
\includegraphics[width=3.8cm,clip]{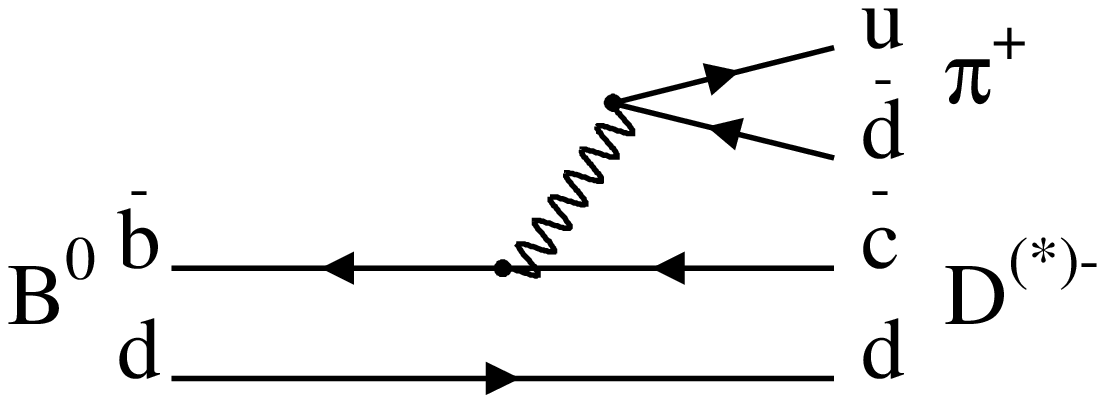}
(a) CFD\\
\end{center}
\end{minipage}
\begin{minipage}{0.2cm}
\hfill
\end{minipage}
\begin{minipage}{4.0cm}
\begin{center}
\includegraphics[width=3.8cm,clip]{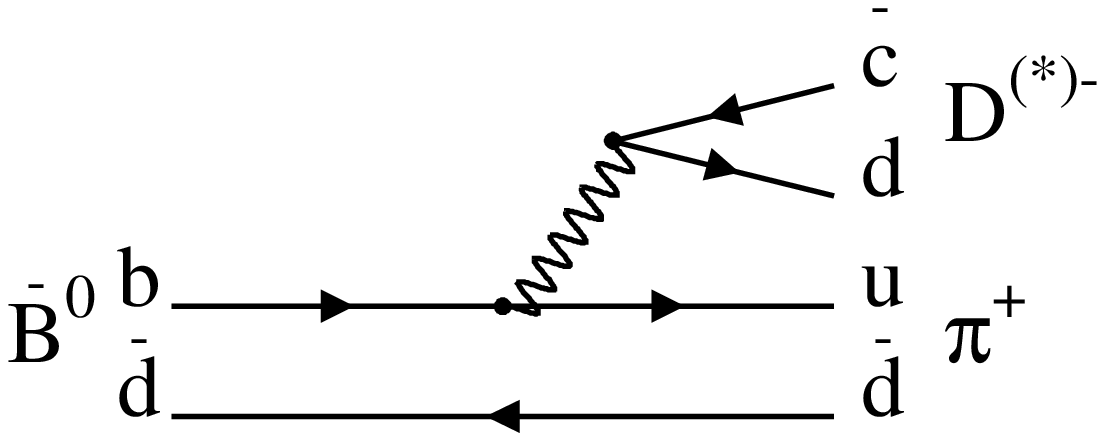}
(b) DCSD\\
\end{center}
\end{minipage}
\caption{
  Contributions to $B^0  \to D^{(*)-}\pi^+$ can come
  either (a) from CFD or (b) from mixing followed by DCSD.
}
\label{diagram}
\end{figure}

There are two ways for a state which is initially $B^0$ 
to be found as $D^{(*)-} \pi^+$ at a later time $t$. 
It can occur either directly through a Cabbibo-favoured decay (CFD) 
or through mixing followed by doubly-Cabbibo-suppressed decay (DCSD), 
as shown  in Fig.~\ref{diagram}. Interference of the two processes 
introduces the term containing $\phi_3$ to the time dependent decay rates, 
 which are given by~\cite{abe,fleischer}  
\begin{eqnarray}
  P(B^{0} \rightarrow D^{(*)+} \pi^-)&=& c \left[ 
    1 - \cos \Delta m t - 2{\Im} \bar{\rho} \sin \Delta m t 
  \right]  
  \nonumber \\
  P(B^{0} \rightarrow D^{(*)-} \pi^+)&=& c \left[
    1 + \cos \Delta m t + 2{\Im} {\rho} \sin \Delta m t 
  \right]  
  \nonumber \\
  P(\bar B^{0} \rightarrow D^{(*)+} \pi^-)&=& c \left[ 
    1 + \cos \Delta m t + 2{\Im} \bar{\rho} \sin \Delta m t 
   \right]  
  \nonumber \\
  P(\bar B^{0} \rightarrow D^{(*)-} \pi^+)&=& c \left[
    1 - \cos \Delta m t - 2{\Im} {\rho} \sin \Delta m t 
   \right] \nonumber \\   \label{eq:evol}
\end{eqnarray}
where $c = (e^{-t/\tau_{B^0}})/2\tau_{B^0}$ with $\tau_{B^0}$ 
denoting the lifetime of the neutral $B$ meson and 
$\Delta m$ is the $B^0$-$\bar{B^0}$ mixing parameter. 
The $\rho$ and $\bar{\rho}$ are defined as 
$\rho = (q/p)({\cal A}(\bar{B^0} \to D^{(*)-} \pi^+)/
              {\cal A}({B^0} \to D^{(*)-} \pi^+))$ and 
$\bar{\rho} = (p/q)({\cal A}({B^0} \to D^{(*)+} \pi^-)/
              {\cal A}(\bar{B^0} \to D^{(*)+} \pi^-))$, where 
$p$ and $q$ relate the mass eigenstates to the flavour eigenstates in the 
neutral $B$ meson system~\cite{abe}. They 
lead to CP violating terms  
${\Im}{\rho} = - (-1)^L R \sin(2\phi_1+\phi_3 - \delta)$ and 
$ {\Im} \bar{\rho} =  (-1)^L R \sin(2\phi_1+\phi_3 + \delta)$,  
where 
$R$ is the ratio of the magnitudes of the DCSD and CFD amplitudes
(here the magnitudes of both the CFD and DCSD amplitudes are assumed 
to be same for $B^0$ and $\bar{B}^0$ decays), $\delta$ is the strong phase 
difference between DCSD and CFD, and $L$ is the angular momentum of 
the final state (1 for $D^* \pi$ and 0 for $D\pi$). 
$R$ and $\delta$ are not necessarily the same for 
$D^* \pi$ and $D \pi$ final states, 
and are denoted with subscripts, $D^* \pi$ and $D \pi$, in what follows. 

This study uses a $140\,\mathrm{fb}^{-1}$ data sample, 
which contains 152 million 
$B \bar B$ 
events, collected with the Belle detector~\cite{Belle} 
at the KEKB collider~\cite{KEKB}.  
The selection of hadronic events is described elsewhere~\cite{hadsel}. 

For the $\bar{B^0} \to D^{*+} \pi^-$ event selection, 
we use the decay chain $D^{*+}\rightarrow D^0 \pi^+$, 
and  $D^0 \to K^- \pi^+,~K^-\pi^+\pi^0$ or $K^-\pi^+\pi^+\pi^-$
(charge conjugate modes are implied throughout this paper).  
For the $\bar{B^0} \to D^+ \pi^-$ event selection,
we use $D^+ \to K^-\pi^+\pi^+$ decays. 
Charged tracks except the slow $\pi^+$ in the $D^{*+} \to D^0 \pi^+$
decay are required to have a minimum of one hit
(two hits) in the $r$-$\phi$ ($z$) plane of the vertex detector
in order to allow precise vertex determination.
To separate kaons from pions, we form a likelihood for each track, 
$\mathcal{L}_{K(\pi)}$. 
The kaon likelihood ratio, 
$P(K/\pi) = \mathcal{L}_K /(\mathcal{L}_K + \mathcal{L}_\pi)$, 
has values between 0 (likely to be a pion) and 1 (likely to be a kaon).
We require charged kaons to satisfy $P(K/\pi)> 0.3$. 
No such requirement is imposed to select charged pions coming from $D$ decays.

For $D^0$ selection, 
the invariant mass of the daughter particles is required to be within 
$\pm 16.5\,\mathrm{MeV}/c^2$, 
$\pm 24.0\,\mathrm{MeV}/c^2$, 
and $\pm 13.5\,\mathrm{MeV}/c^2$ 
of the nominal $D^0$ mass, 
for $K^-\pi^+$, $K^-\pi^+\pi^0$, and $K^-\pi^+\pi^+\pi^-$ modes, respectively. 
These intervals correspond to $\pm 3\sigma$, 
where $\sigma$ is the Monte Carlo determined invariant mass resolution. 
For the $D^+$, 
the invariant mass is required to be within 
$\pm 12.5\,\mathrm{MeV}/c^2$ of the nominal $D^+$ mass. 
For the $D^0 \to K^- \pi^+ \pi^0$ reconstruction, 
we further require the $\pi^0$ momentum to be greater than 
$200\,\mathrm{MeV}/c$ in the $\Upsilon(4S)$ rest frame, and  
the ratio of the second to zeroth Fox-Wolfram moments~\cite{fw}
$R_2$ to be less than $0.55$.
We require $R_2 < 0.5$ for $D^+ \to K^-\pi^+\pi^+$.  
We use a mass- and vertex-constrained fit for $D^0$ and a vertex-constrained 
fit for $D^+$.

The $D^{*+}$ is reconstructed by combining $D^0$ candidates with a slow 
$\pi^+$.
Here, slow pions are required to have momentum  
less than $300\,\mathrm{MeV}/c$ in the $\Upsilon(4S)$ rest frame. 
The $D^*$ candidates are required to have a mass difference
$\Delta M \equiv M_{D^0\pi}-M_{D^0}$ within 
$\pm 7\,\mathrm{MeV}/c^2$, $\pm 2\,\mathrm{MeV}/c^2$, or 
$\pm 4\,\mathrm{MeV}/c^2$ of the nominal value, 
for the 
$K^-\pi^+$, $K^-\pi^+\pi^0$, and $K^-\pi^+\pi^+\pi^-$ modes respectively. 

We reconstruct $B$ candidates by combining the $D^{(*)+}$ candidate with 
a $\pi^-$ candidate satisfying $P(K/\pi) < 0.8$. 
We identify $B$ decays based on requirements on the energy difference 
$\Delta E \equiv \sum_i E_i - E_{\rm{beam}}$ and the  
beam-energy constrained mass 
$M_{\rm{bc}} \equiv \sqrt{E_{\rm{beam}}^2 - (\sum_i \vec{p}_i)^2}$, 
where $E_{\rm{beam}}$ is the beam energy, $\vec{p}_i$ and $E_i$ are 
the momenta and energies of the daughters of the reconstructed $B$ 
meson candidate, all in the $\Upsilon(4S)$ rest frame. 
If more than one $B$ candidate is found in the same event, 
we select the one with best $D$ vertex quality.
We define a signal region in the 
$\Delta E$-$M_{\mathrm{bc}}$ plane of 
$5.27\,\mathrm{GeV}/c^2 < M_{\rm bc} < 5.29\,\mathrm{GeV}/c^2$ and 
$\left| \Delta E \right| < 0.045\,\mathrm{GeV}$, 
corresponding to about $\pm 3\sigma$ of both quantities.
For the determination of background parameters, 
we use events in a sideband region defined by 
$M_{\rm bc} > 5.2\,\mathrm{GeV}/c^2$ and
$-0.14\,\mathrm{GeV} < \Delta E  < 0.20\,\mathrm{GeV}$, 
excluding the signal region. 

Charged leptons, pions, and kaons that are not associated with the
reconstructed $D^{(*)} \pi$ decays are used to identify the flavour of
the accompanying $B$ meson. 
The algorithm~\cite{belle_sin2phi1} leads to two parameters, $q$ and $r$, 
where $q=+1$ indicates $\bar b$ hence
$B^0$ and $q=-1$ indicates $b$ hence $\bar{B}^0$. The parameter $r$ is an
event-by-event dilution factor ranging from $r=0$ for no flavour
discrimination to $r=1$ for unambiguous flavour assignment.  
More than 99.5\% of the events are assigned non-zero values of $r$.       

The decay vertices of the $B \to D^{(*)} \pi$ are fitted using 
the momentum vectors of the $D$ and $\pi$ (except the slow $\pi$ 
from $D^{*}$ decay) and a requirement that they are consistent with the 
interaction region profile. 
For the decay vertices of the tagging $B$ meson, the remaining 
well reconstructed tracks in the event are used. 
Tracks that are consistent with $K^0_S$ decay are rejected. 
The proper-time difference between the fully reconstructed and 
the associated $B$ decay is calculated as  
$\Delta t = (z_{\mathrm{rec}}-z_{\mathrm{tag}})/c\beta \gamma$, 
where $z_\mathrm{rec}$ and $z_\mathrm{tag}$ are the $z$ coordinates of the two
$B$ decay vertices and $\beta \gamma = 0.425$ is the
Lorentz boost factor at KEKB. 
After application of the event selection criteria and the  requirement that 
both $B$'s have well defined vertices and 
$\left| \Delta t \right| < 70 \ {\rm ps}$ ($\sim 45\, \tau_{B^0}$),  
7763 and 9351 events remain as the $D^* \pi$ and $D\pi$ candidates, 
respectively. The signal fractions of the samples, which vary for 
different $r$ bins, are 96\% for $D^* \pi$ and 91\% for $D \pi$.  

Unbinned maximum likelihood fits to the four time dependent decay
rates are performed to extract
${\Im} \rho$ and ${\Im} \bar{\rho}$. 
We minimize $-2 \sum_{i} \ln L_i$ where the likelihood for
the $i$-th event is given by
\begin{eqnarray*} 
 L_i &=& (1 - f_{\rm ol})
  \left[
    f_{\rm sig} P_{\rm sig}\otimes R_{\rm sig} 
    + (1-f_{\rm sig}) P_{\rm bkg}\otimes R_{\rm bkg}
  \right]\\
 && + f_{\rm ol} P_{\rm ol}.
\end{eqnarray*}
The signal fraction $f_{\rm sig}$ is determined from 
the ($\Delta E$, $M_{\rm{bc}}$) value of each event. 
The signal distribution is the product of 
the sum of two Gaussian in $\Delta E$ and a Gaussian in $M_{\rm bc}$;
that for the background is the product of a first order polynomial 
in $\Delta E$ and an ARGUS function~\cite{argus} in $M_{\rm bc}$.

The $\Delta t$ distribution is modeled by a core distribution convolved 
with resolutions.  
A small number of events have poorly reconstructed vertices resulting 
in a very broad distribution $\Delta t$. 
We account for the contributions from these ``outliers'' by
adding a Gaussian component $P_{\rm ol}$ with a width and fraction 
determined from the $B$ lifetime analysis~\cite{tajima}.
The $\Delta t$ resolution, denoted by $R_{\rm sig}$ and $R_{\rm bkg}$
for the signal and background, is determined on an event-by-event basis,
using the estimated uncertainties on the $z$ vertex positions
~\cite{response}.  

The signal $\Delta t$ distributions are given by
\begin{eqnarray}
&&P_{\rm sig}(q=-1, D^{(*)\pm} \pi^{\mp})  \nonumber \\  
 && = (1- w_-) P(B^0 \to D^{(*)\pm} \pi^{\mp}) + 
       w_+  P(\bar{B^0} \to D^{(*)\pm} \pi^{\mp}) \nonumber \\
&&P_{\rm sig}(q=+1, D^{(*)\pm} \pi^{\mp})  \nonumber \\  
 && = (1- w_+) P(\bar{B^0} \to D^{(*)\pm} \pi^{\mp}) + 
       w_-  P({B^0} \to D^{(*)\pm} \pi^{\mp}) \nonumber \\
   \label{eq:pdf}
\end{eqnarray}
where   
$w_-$ and $w_+$ are wrong tag fractions for the $q=-1$ and $q=+1$ samples, 
respectively. $P$'s are given by Eq.~\ref{eq:evol} with $t$ and $c$
replaced by $\Delta t$ and $(e^{-|\Delta t|/{\tau_{B^0}}})/4\tau_{B^0}$, 
respectively. 

The background $\Delta t$ distribution is parameterized as a sum of 
a $\delta$-function
component and an exponential component with a lifetime $\tau_{\rm bkg}$
\begin{equation}
  P_{\rm bkg} = f_{\rm bkg}^{\delta} 
  \delta(\Delta t -\mu_{\rm bkg}^{\delta}) + 
  \frac{(1-f_{\rm bkg}^{\delta})}{2\tau_{\rm bkg}}
  e^{-\left| \Delta  t -\mu_{\rm bkg}^\tau \right|/{\tau_{\rm bkg}}}
  \nonumber
\end{equation}
where $f_{\rm bkg}^{\delta}$ is the fraction of events 
contained in the $\delta$-function, 
and $\mu_{\rm bkg}^{\delta}$ and $\mu_{\rm bkg}^\tau$
are the mean values of $\left| \Delta t \right|$ in the 
$\delta$-function and exponential components, respectively.

While the tagging side should have no asymmetry if the flavour is tagged by 
primary leptons, it is possible to introduce a small 
asymmetry when daughter particles of hadronic 
decays such as $D^{(*)} \pi$ are used for the flavour tagging,
due to the same CP violating effect, which is the subject of 
this paper ~\cite{tagsidecpv}. This effect is taken into account by 
replacing the coefficients of $\sin \Delta m t$ in Eqs.~\ref{eq:evol} by 
${\Im} \bar{\rho} - {\Im} \bar{\rho}^{\prime}$, 
${\Im} {\rho}     - {\Im} \bar{\rho}^{\prime}$, 
${\Im} \bar{\rho} - {\Im} {\rho}^{\prime}$, and  
${\Im} {\rho}     - {\Im} {\rho}^{\prime}$, respectively.  
Here the ${\Im} \rho^{\prime}$ and ${\Im} \bar{\rho^{\prime}}$ 
represent the CP violating effect due to the presence of 
$B^0 \to \bar{D} X$ and $B^0 \to D X$ amplitudes in the flavour tagging side. 
Note that unlike the ${\Im} \rho$ and ${\Im} \bar{\rho}$, which are 
rigorously defined in terms of 
$B^0 \to D^{(*)\mp} \pi^{\pm}$ and $\bar{B^0} \to D^{(*)\pm} \pi^{\mp}$ 
amplitudes, ${\Im} \rho^{\prime}$ and  ${\Im} \bar{\rho}^{\prime}$ 
are effective quantities that include effects of the fraction of $B \to D X$ 
components in the tagging $B$ decays and all experimental effects of 
subsequent behaviour of $D$ mesons. 
Therefore, these quantities must be determined experimentally. 

The values of ${\Im} {\rho^{\prime}}$ and 
${\Im}\bar{\rho^{\prime}}$ are determined in each of six $r$ bins   
by fitting the $\Delta t$ distributions 
of a $D^*l\nu$ control sample~\cite{dstarlnu} 
using the signal distributions of Eq.~\ref{eq:pdf} and setting 
${\Im} {\rho}$ and ${\Im} \bar{\rho}$ to zero. 
Since the $D^*l\nu$ final states have specific flavour, any observable 
asymmetry must originate from the tagging side.  
The results for the combined $r$ bins are 
$2{\Im} {\rho^{\prime}} = 0.038 \pm 0.014(\mathrm{stat}) \pm 0.005
(\mathrm{sys})$ and 
$2{\Im} \bar{\rho^{\prime}} =  0.002 \pm 0.014(\mathrm{stat}) 
\pm 0.009(\mathrm{sys})$.

The procedures for $\Delta t$ determination and flavour tagging are
tested by extracting $\tau_{B^0}$ and $\Delta m$. 
When all four signal categories in Eq.~\ref{eq:evol} are combined, 
the signal $\Delta t$ distribution reduces to an exponential lifetime 
distribution.  
We obtain 
$\tau_{B^0} = 1.583 \pm 0.029 \ {\rm ps}$ ($1.575 \pm 0.032 \ {\rm ps}$) 
for the $D^* \pi$ ($D \pi$) samples, 
in good agreement with the world average 
$(1.542 \pm 0.016 \ {\rm ps})$~\cite{PDG}.
Combining the two CFD-dominant modes and the two mixing-dominant modes 
and ignoring the CP violating terms,  
the asymmetry behaves as $\cos \Delta m \Delta t$.
We obtain 
$\Delta m = 0.490 \pm 0.015 \ {\rm ps}^{-1}$ 
($0.483 \pm 0.014 \ {\rm ps}^{-1}$) for the $D^* \pi$ ($D \pi$) samples,  
also in good agreement with the world average 
$(0.489 \pm 0.008 \ {\rm ps}^{-1})$~\cite{PDG}. 
The same fits also provide wrong tag fractions $w_-$ and $w_+$ 
in each $r$ bin for both $D^* \pi$ and $D \pi$ data samples. 
The errors of our results are statistical only.

\begin{figure}[!thb]
  \begin{center}
    \begin{tabular}{cc}
      \begin{minipage}{4.3cm}
        \begin{center} 
          \includegraphics[width=4.3cm,clip]{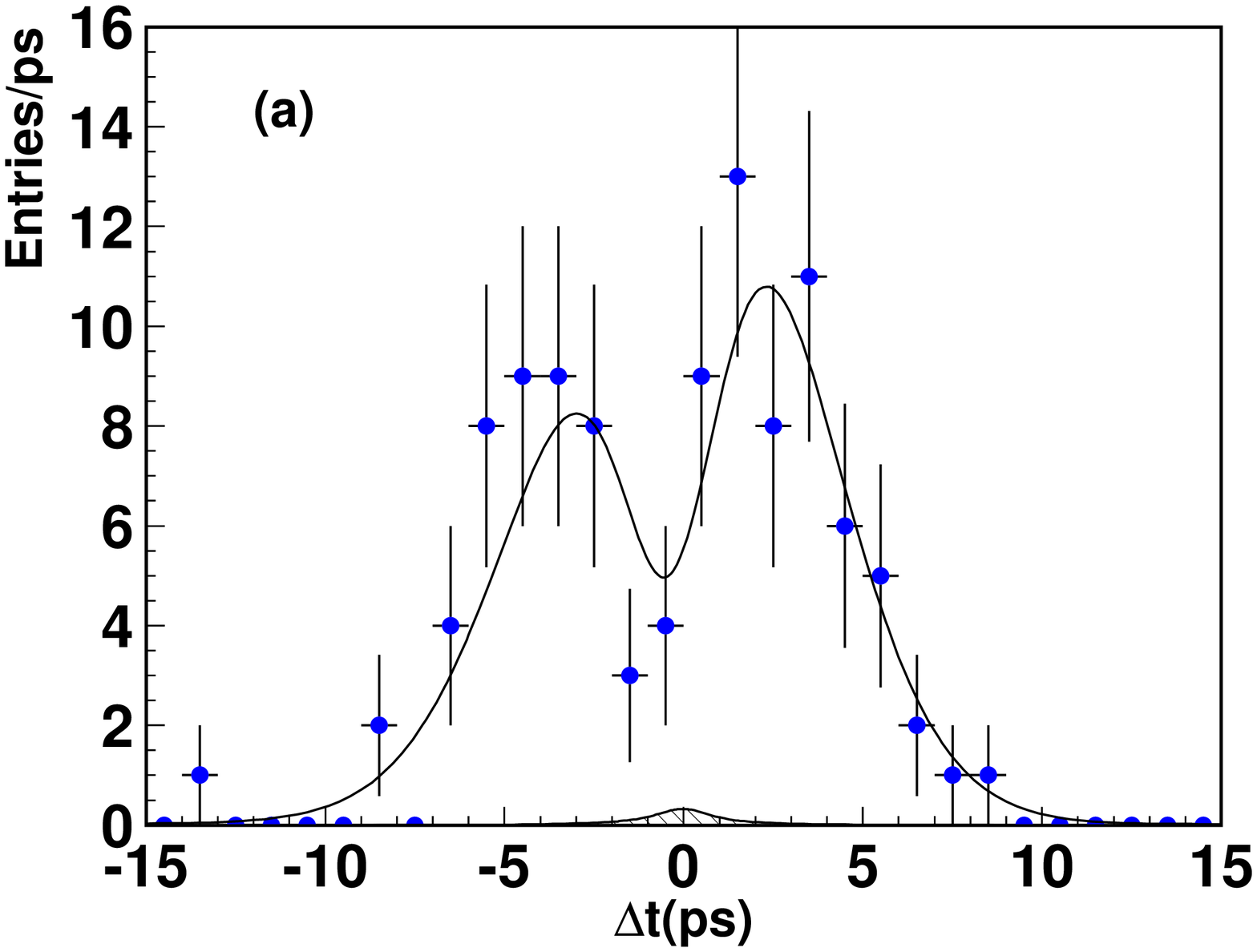} 
        \end{center}
      \end{minipage}
      &
      \begin{minipage}{4.3cm}
        \begin{center} 
          \includegraphics[width=4.3cm,clip]{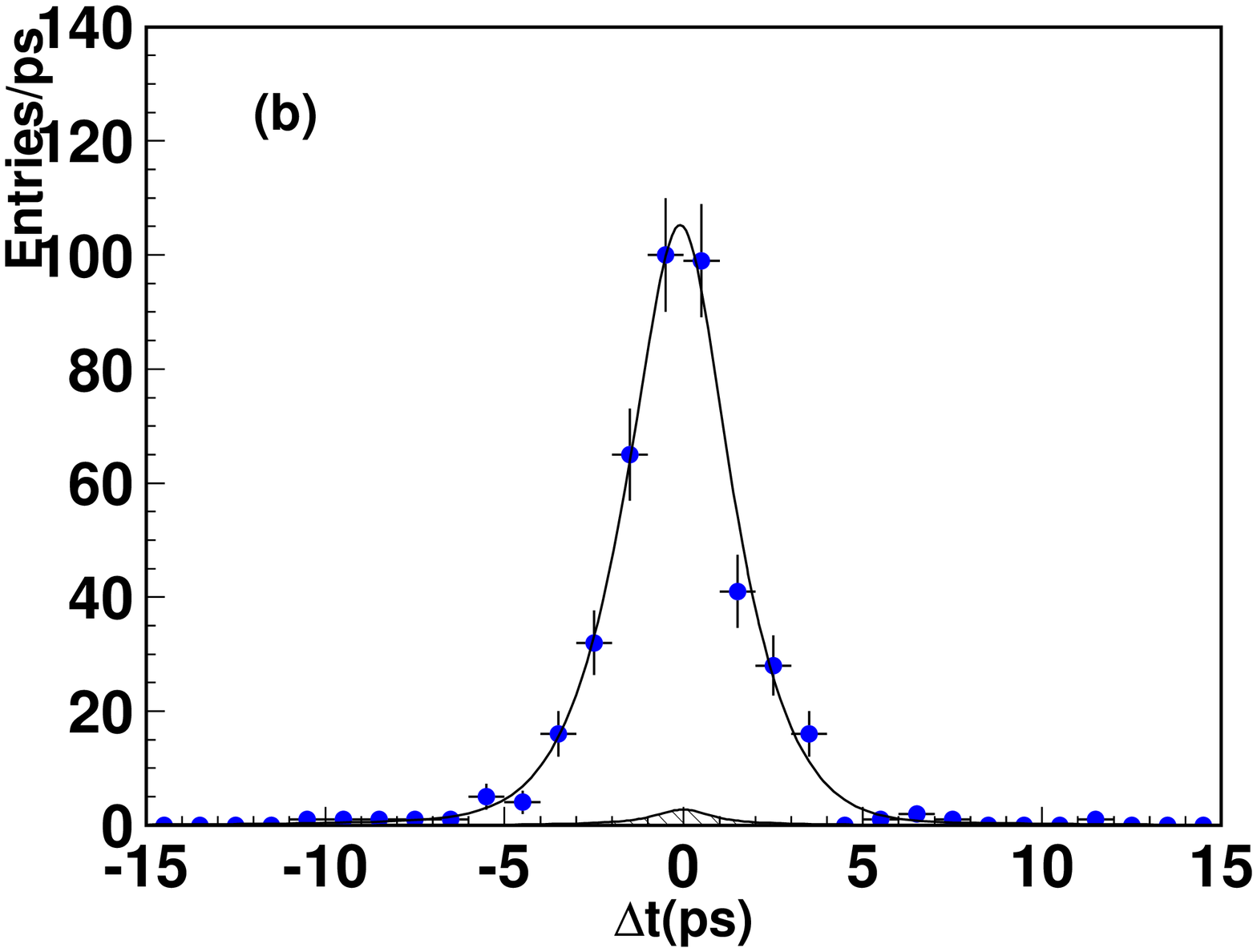} 
        \end{center}
      \end{minipage}
      \\ 
      \begin{minipage}{4.3cm}
        \begin{center} 
           \includegraphics[width=4.3cm,clip]{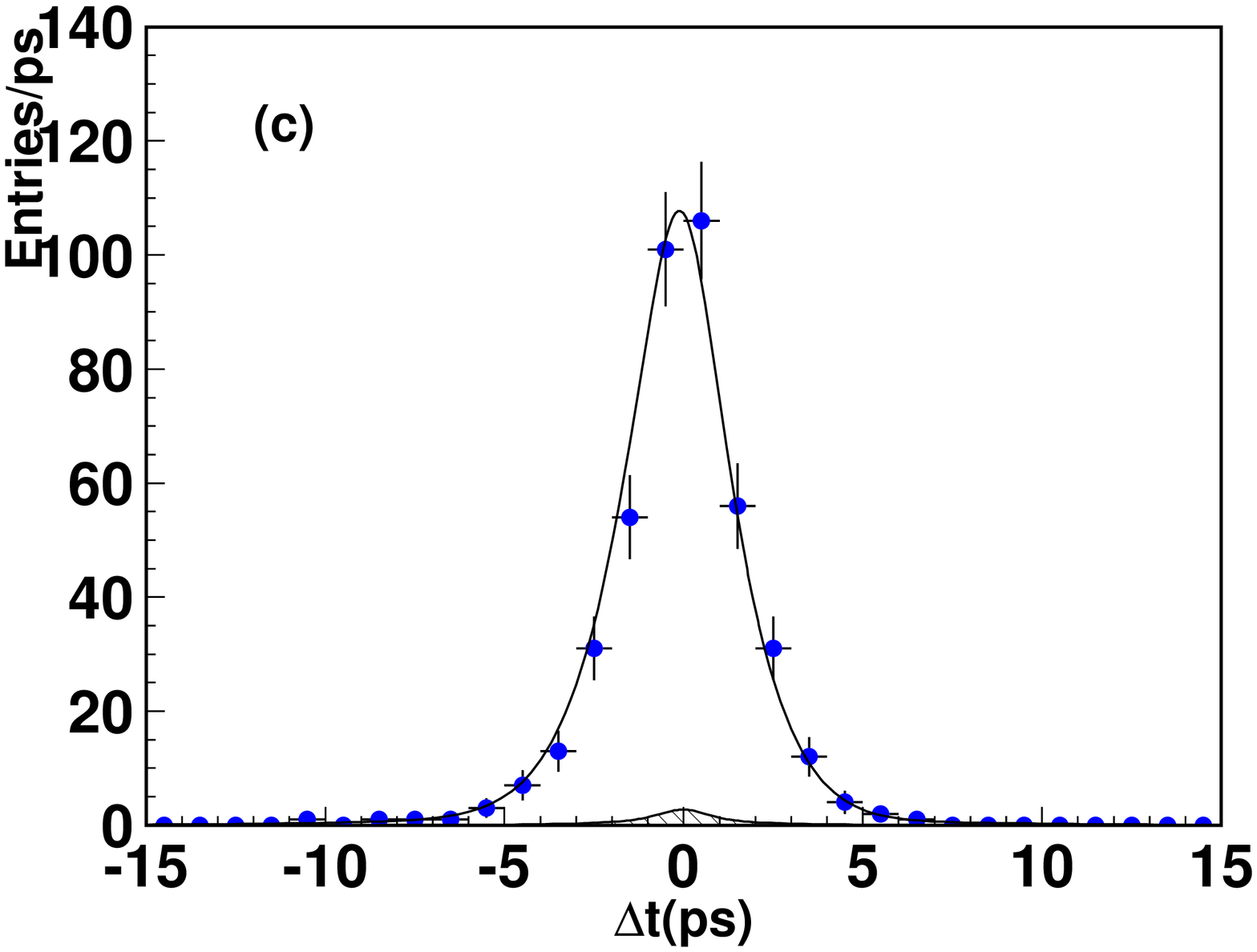} 
        \end{center}
      \end{minipage}
      &
      \begin{minipage}{4.3cm}
        \begin{center} 
      \includegraphics[width=4.3cm,clip]{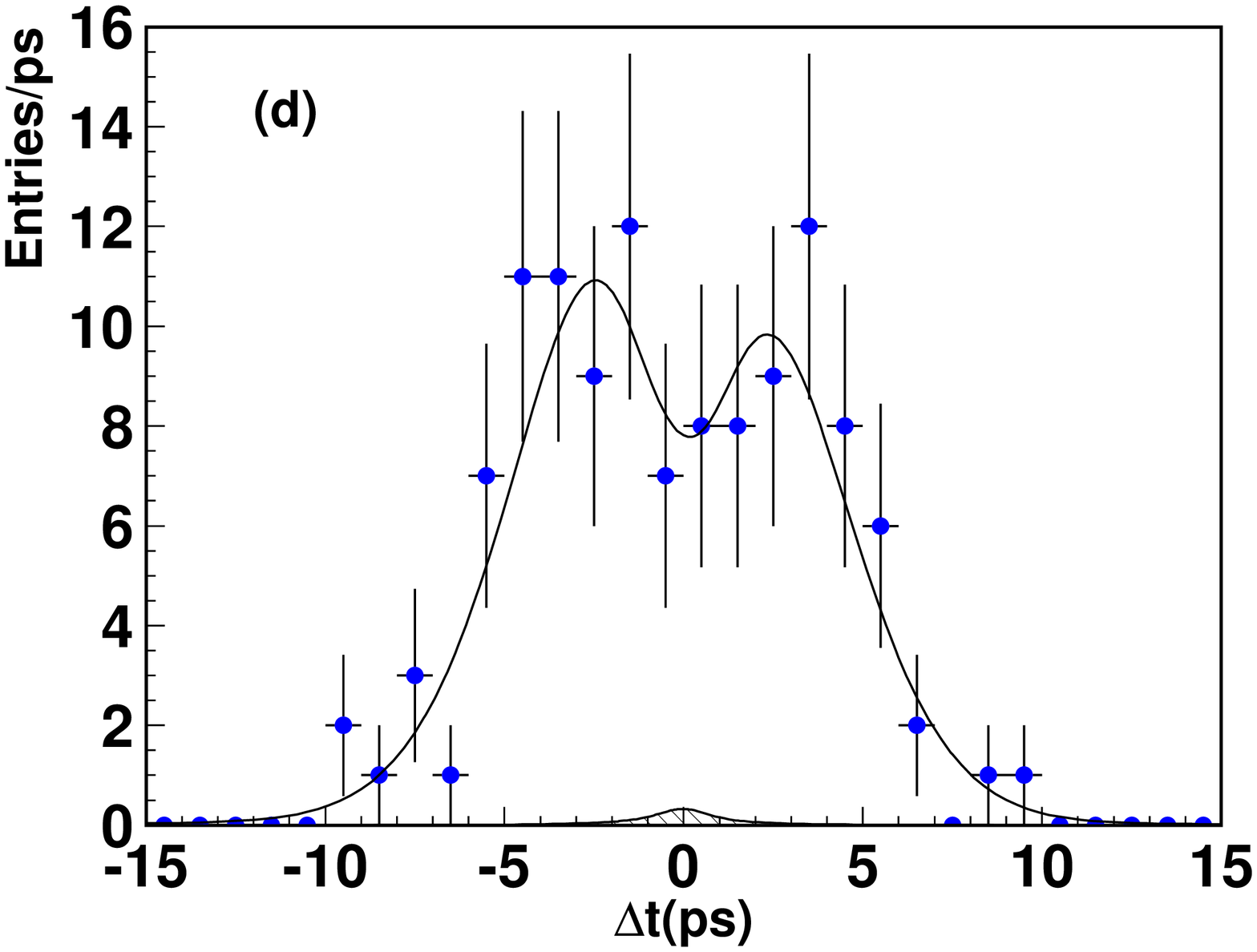} 
        \end{center}
      \end{minipage}
    \end{tabular}
  \end{center}
\vspace*{-5mm}  
\caption{
    \label{dt_dstpi_r6}
    $\Delta t$ distributions for the $D^* \pi$ data in the 
    $0.875 < r < 1.000$ flavour tagging quality bin.
    (a) $B^0 \to D^{*+} \pi^-$, 
    (b) $B^0 \to D^{*-} \pi^+$, 
    (c) $\bar{B}^0 \to D^{*+} \pi^-$, 
    (d) $\bar{B}^0 \to D^{*-} \pi^+$.
    Curves show the fit results with the entire event sample, 
    hatched regions indicate the backgrounds.
  }
  \begin{center}
    \begin{tabular}{cc}
       \begin{minipage}{4.3cm}
        \begin{center} 
          \includegraphics[width=4.3cm,clip]{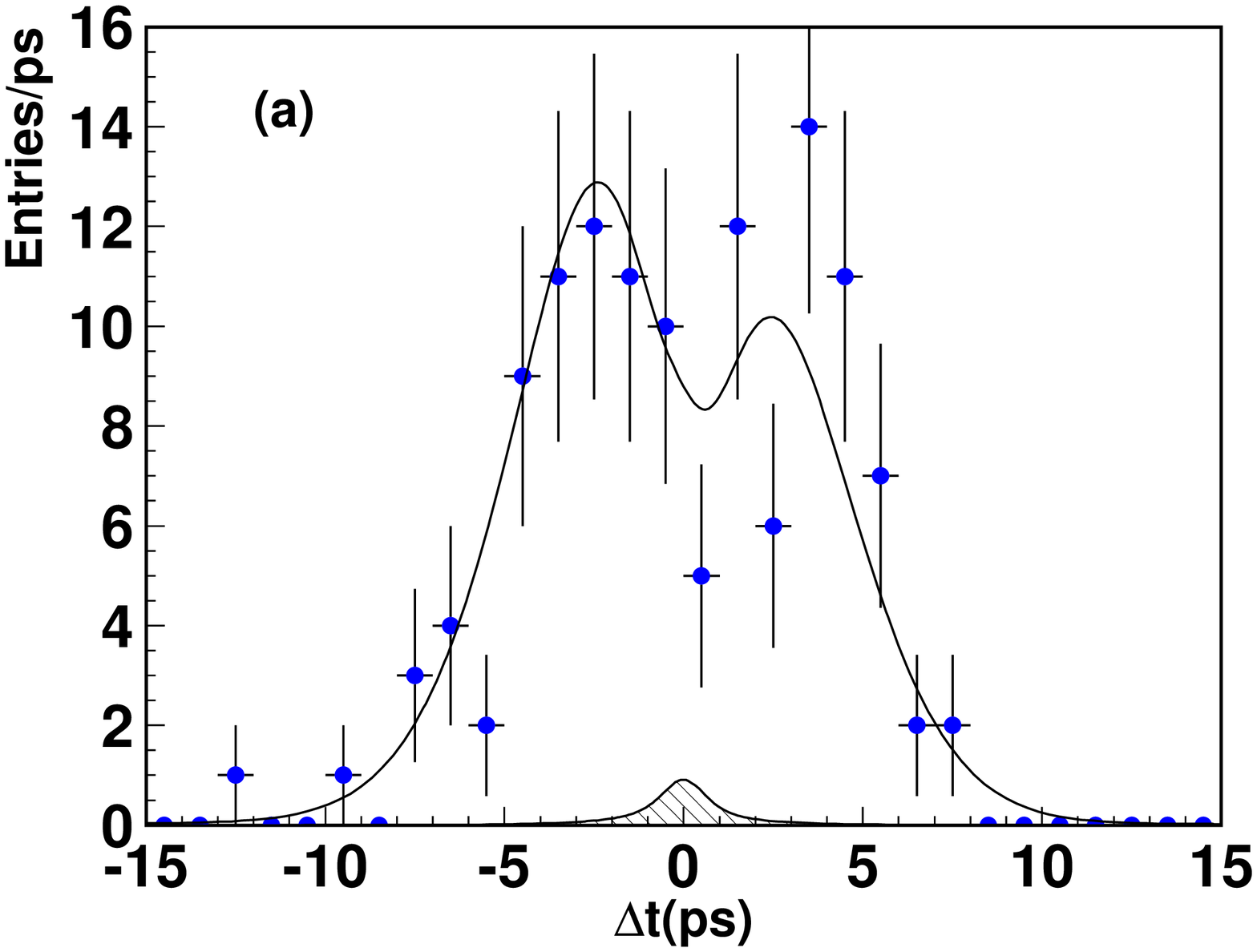} 
        \end{center}
      \end{minipage}
      &
      \begin{minipage}{4.3cm}
        \begin{center} 
           \includegraphics[width=4.3cm,clip]{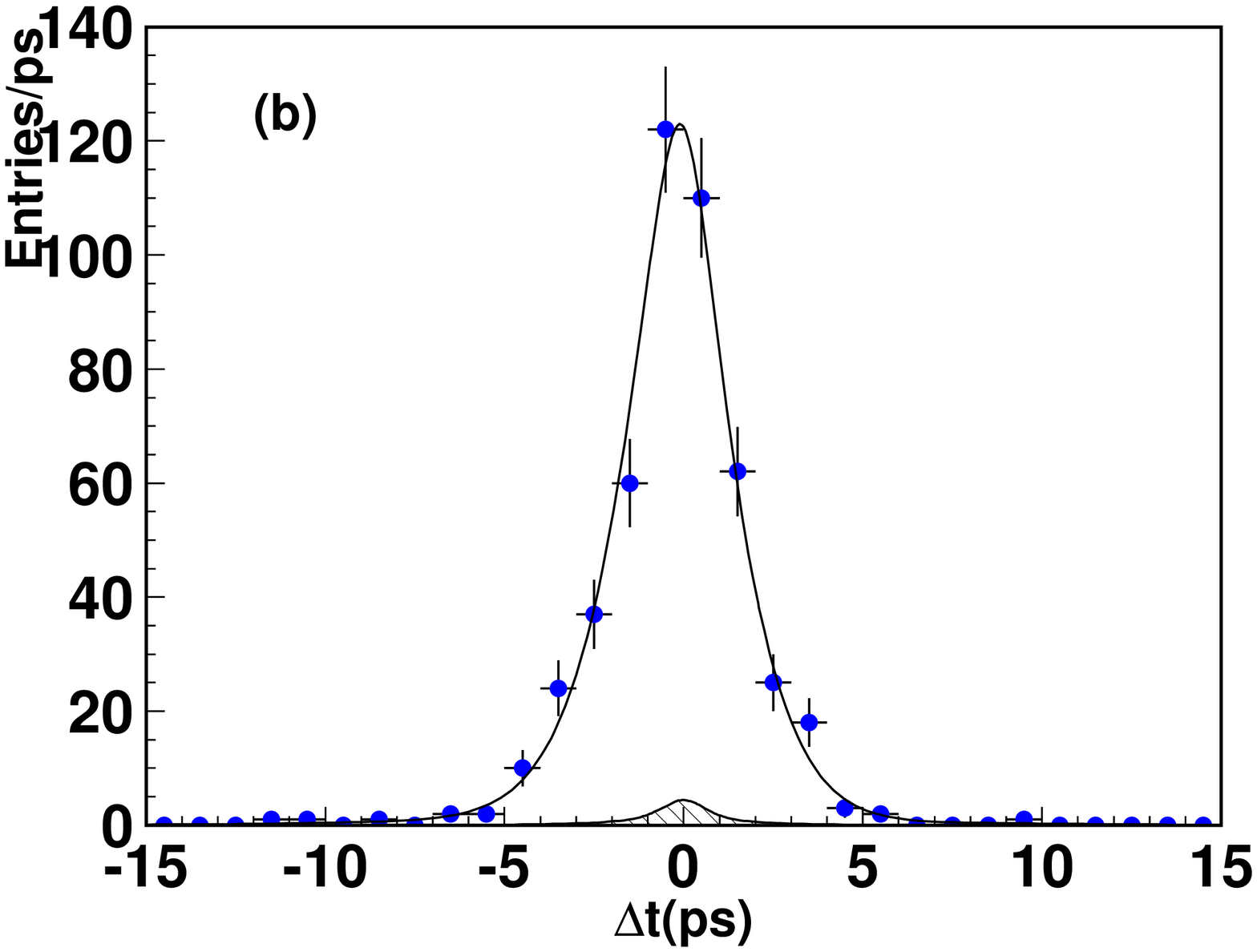} 
        \end{center}
      \end{minipage}
      \\
      \begin{minipage}{4.3cm}
        \begin{center} 
          \includegraphics[width=4.3cm,clip]{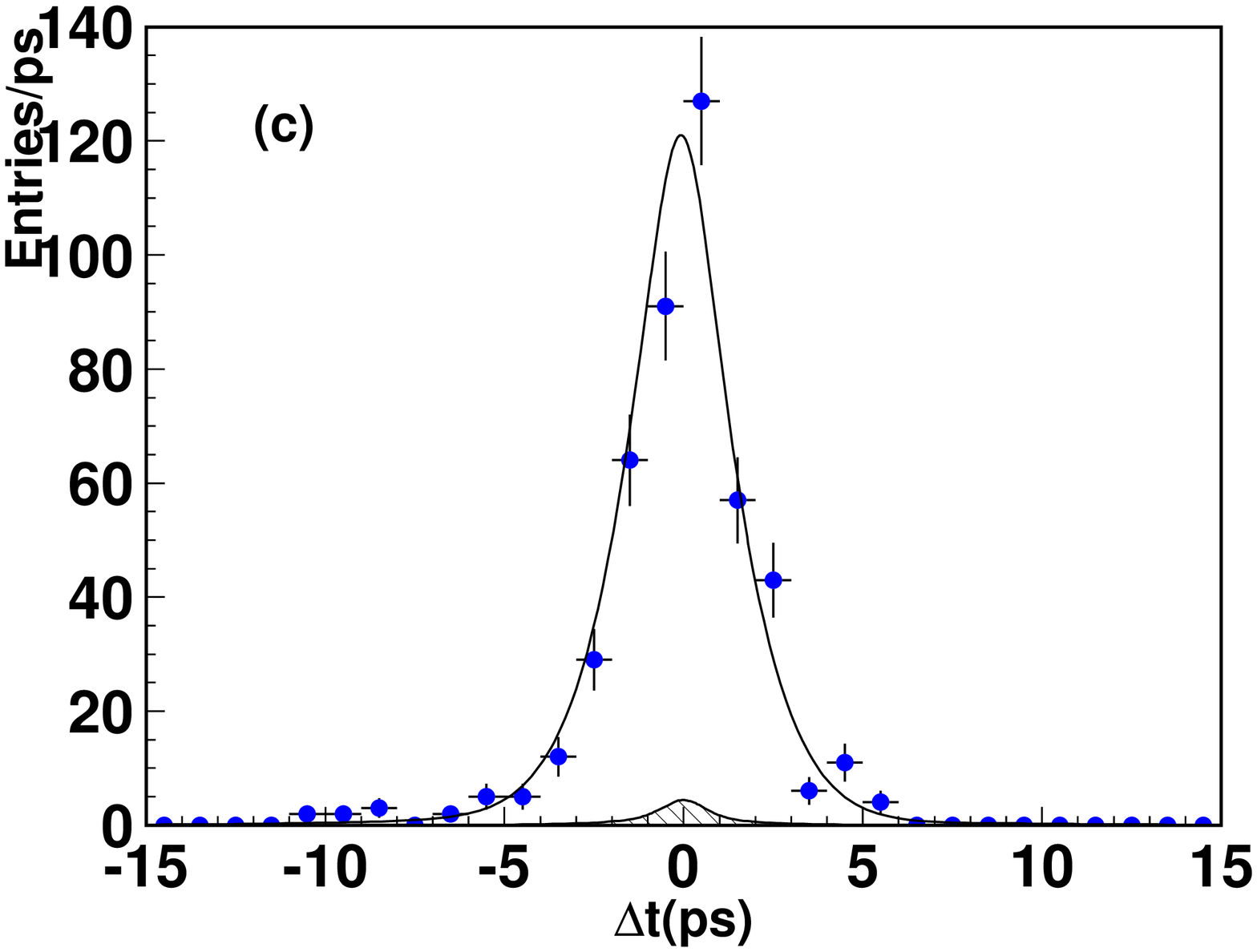} 
        \end{center}
      \end{minipage}
      &
      \begin{minipage}{4.3cm}
        \begin{center} 
           \includegraphics[width=4.3cm,clip]{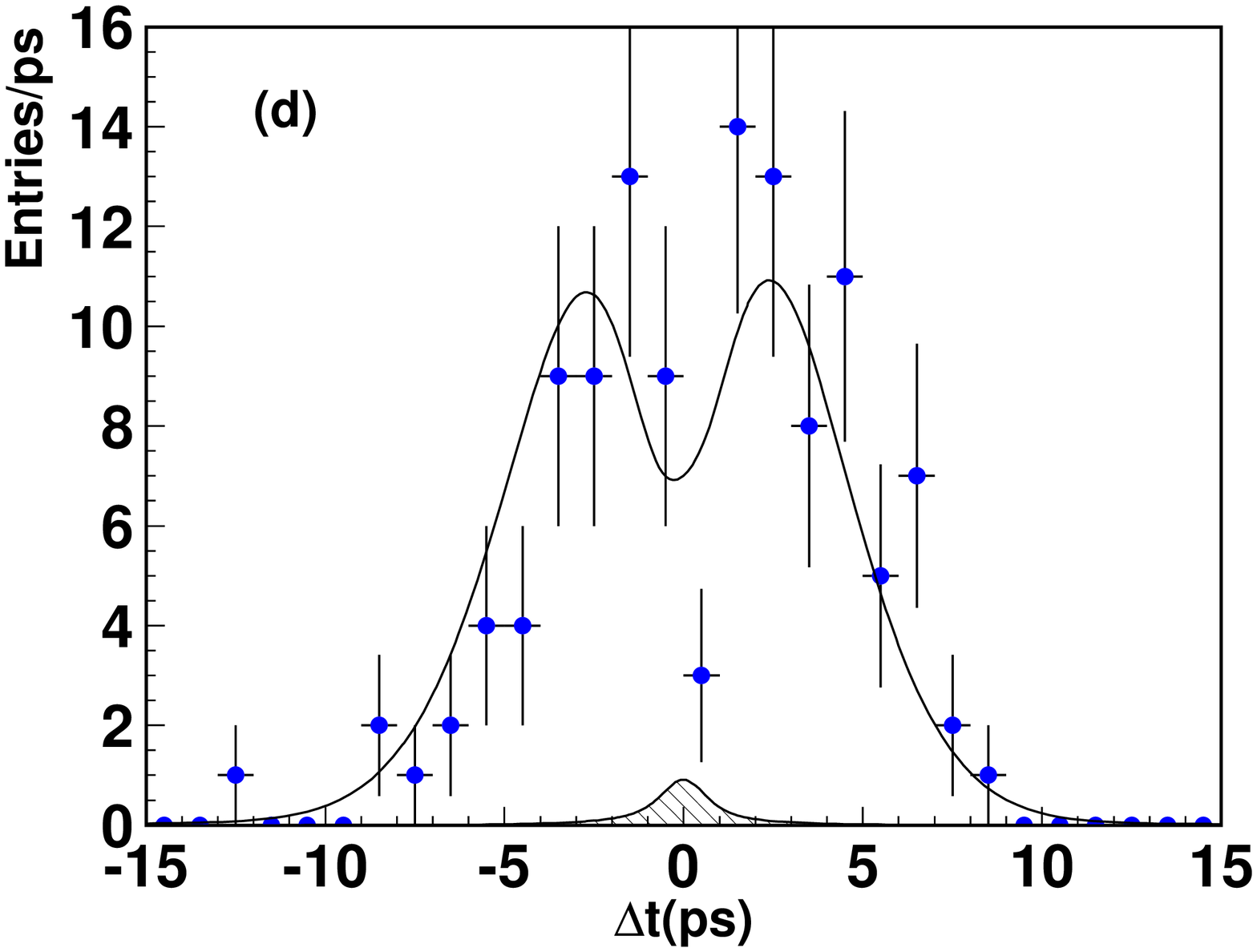} 
        \end{center}
      \end{minipage}
    \end{tabular}
  \end{center}
  \vspace*{-5mm}
  \caption{
    \label{dt_dpi_r6}
    $\Delta t$ distributions for the $D \pi$ events in the 
    $0.875 < r < 1.000$ flavour tagging quality bin.
    (a) $B^0 \to D^{+} \pi^-$, 
    (b) $B^0 \to D^{-} \pi^+$, 
    (c) $\bar{B}^0 \to D^{+} \pi^-$, 
    (d) $\bar{B}^0 \to D^{-} \pi^+$.
    Curves show the fit results with the entire event sample, 
hatched regions indicate the backgrounds.
  }
\end{figure}

We then perform fits to determine the ${\Im} \rho$ and 
${\Im} \bar{\rho}$  
by fixing $\tau_{B^0}$ and  $\Delta m_d$  to the world 
average values and using $w_-$, $w_+$, 
${\Im} \rho^{\prime}$, and ${\Im} \bar{\rho^{\prime}}$ 
for each $r$ bin, as obtained from the above fits. The results are 
$2{\Im} {\rho}_{D^* \pi} =  0.011 \pm 0.057$, 
$2{\Im} {\bar{\rho}}_{D^* \pi} = -0.109 \pm 0.057$, 
$2{\Im} {\rho}_{D \pi} =  -0.037 \pm 0.052$, and 
$2{\Im} {\bar{\rho}}_{D \pi} = 0.087 \pm 0.054$. 
The errors are statistical only. 
The $\Delta t$ distributions for the subsamples having the best quality 
flavour tagging ($0.875 < r < 1.000$) are shown in 
Fig.~\ref{dt_dstpi_r6} for the $D^* \pi$ and in
Fig.~\ref{dt_dpi_r6} for the $D \pi$ samples, respectively. 

The systematic errors come from 
i) the uncertainties of parameters which are constrained in the fit,
including $\Delta t$ resolution parameters, 
background parameters, wrong tag fractions, and physics parameters;
ii) uncertainties of the tagging side asymmetries; 
iii) fit biases induced by the vertexing and other unknown factors. 
For item i), we repeat the fits 
varying each parameter value by $\pm 1\sigma$.  
To estimate item ii), 
we repeat the fits by varying the ${\Im} {\rho^{\prime}}$ and 
${\Im} \bar{\rho^{\prime}}$ by their errors. 
Errors are not explicitly assigned for item iii), since they are
included in the errors of $\Im \rho^{\prime}$ and $\Im \bar{\rho}^{\prime}$ 
from the $D^* l \nu$ control sample fit (item ii).  
Table~\ref{syserror} summarizes the systematic errors. 
\begin{table}[htb]
\caption{
  Systematic errors in the $2R \sin (2\phi_1 + \phi_3 \pm \delta)$ extractions.
}
\label{syserror}
\begin{tabular}{lcc}
\hline \hline
Sources   
 &  $D^* \pi$  &  $D \pi$ \\
\hline
Signal $\Delta t$ resolution       & 0.014    & 0.013   \\
Background $\Delta t$ shape        & 0.001    & 0.003   \\
Background fraction                & 0.002    & 0.001   \\
Wrong tag fraction                 & 0.006    & 0.006   \\
Vertexing                          & 0.005    & 0.005   \\
Physics parameters ($\Delta m, \tau_{B^0}$)    
                                   & 0.001    & 0.002   \\
Tagging side asymmetry             & 0.009    & 0.009   \\
\hline
Combined                           & 0.019    & 0.018   \\
\hline \hline
\end{tabular}
\end{table}

We obtain
\begin{eqnarray}
  2 R_{D^* \pi} \sin (2\phi_1 + \phi_3 + \delta_{D^* \pi})&=&
0.109 \pm 0.057 \pm 0.019,
  \nonumber \\
  2 R_{D^* \pi} \sin (2\phi_1 + \phi_3 - \delta_{D^* \pi}) &=&
  0.011 \pm 0.057 \pm 0.019,
  \nonumber \\
  2 R_{D \pi} \sin (2\phi_1 + \phi_3 + \delta_{D \pi}) &=&
  0.087 \pm 0.054 \pm 0.018,
  \nonumber \\
  2 R_{D \pi} \sin (2\phi_1 + \phi_3 - \delta_{D \pi}) &=&
  0.037 \pm 0.052 \pm 0.018.
  \nonumber
\end{eqnarray}
The first and second errors are statistical and systematic. 
At present, the statistical errors are too large to allow any meaningful 
conclusion to be drawn. 
However, it is interesting to consider how the four results can 
be combined using knowledge of $R$ and $\delta$ to improve 
the precision of $\sin (2\phi_1 + \phi_3)$.  
Several methods have been proposed to measure $R$~\cite{dunietz}. 
A method that compares the branching fractions of 
$B^0 \to D_s^{(*)+} \pi^-$ and $B^0 \to D^{(*)-} \pi^+$ and uses 
factorization relation gives $R_{D \pi} = 0.0237 \pm 0.0050$ and 
$R_{D^* \pi} = 0.0180 \pm 0.0067$\cite{restimate}. 
The present errors are too large to conclude that the two $R$ values are 
equal. 
On the other hand, there are solid theoretical grounds for assuming  
$\delta_{D^* \pi}$ and $\delta_{D \pi}$ to be very small and therefore 
equal~\cite{wolfenstein}. However, some argue that there is an 
ambiguity of $180^{\circ}$ between $\delta_{D^* \pi}$ and $\delta_{D \pi}$
~\cite{fleischer}. Assuming $\delta_{D^{(*)} \pi}$ is close to either 
$0^{\circ}$ or $180^{\circ}$, we obtain  
$|2R_{D^* \pi} \sin (2\phi_1 + \phi_3)| = 0.060 \pm 0.040(\mathrm{stat}) 
\pm 0.019(\mathrm{sys})$ and 
$|2R_{D \pi} \sin (2\phi_1 + \phi_3)| = 0.061 \pm 0.037(\mathrm{stat}) 
\pm 0.018(\mathrm{sys})$. 

In summary, we measure the time dependent CP violation parameters 
$2 R \sin (2\phi_1 +\phi_3 \pm \delta)$ for the 
$B^0 (\bar{B^0}) \to D^{(*)\mp} \pi^{\pm}$ decays using 152 million 
$B \bar B$ events. 
Under the assumption of $\delta_{D^{(*)} \pi}$ being close 
to either $0^{\circ}$ or $180^{\circ}$,  
we obtain 
$|2R_{D^* \pi} \sin (2\phi_1 + \phi_3)| = 0.060 \pm 0.040(\mathrm{stat}) 
\pm 0.019(\mathrm{sys})$ and 
$|2R_{D \pi} \sin (2\phi_1 + \phi_3)| = 0.061 \pm 0.037(\mathrm{stat}) 
\pm 0.018(\mathrm{sys})$.

We wish to thank the KEKB accelerator group for the excellent
operation of the KEKB accelerator.
We acknowledge support from the Ministry of Education,
Culture, Sports, Science, and Technology of Japan
and the Japan Society for the Promotion of Science;
the Australian Research Council
and the Australian Department of Education, Science and Training;
the National Science Foundation of China under contract No.~10175071;
the Department of Science and Technology of India;
the BK21 program of the Ministry of Education of Korea
and the CHEP SRC program of the Korea Science and Engineering Foundation;
the Polish State Committee for Scientific Research
under contract No.~2P03B 01324;
the Ministry of Science and Technology of the Russian Federation;
the Ministry of Education, Science and Sport of the Republic of Slovenia;
the National Science Council and the Ministry of Education of Taiwan;
and the U.S. Department of Energy.

\end{document}